\documentclass[journal]{IEEEtran}

\usepackage[colorlinks]{hyperref}
\hypersetup{
    colorlinks,
    linkcolor=black,
    citecolor=black,
    filecolor=magenta,
    urlcolor=cyan,
}

\usepackage{resizegather}

\usepackage{cite}
\usepackage{amsmath,amssymb,amsfonts,mathrsfs}
\usepackage{graphicx}
\usepackage{wrapfig}
\usepackage{mdframed}
\usepackage{textcomp}
\usepackage{textcomp}
\usepackage{tablefootnote}
\usepackage{threeparttable}
\usepackage{caption, subcaption}
\usepackage{bbm}
\usepackage{dsfont}
\usepackage{tikz}
\usepackage{graphicx}
\usepackage{tkz-euclide}
\usepackage{algorithm}
\usepackage{algpseudocode}
\usepackage{booktabs}
\usepackage[subtle]{savetrees}
\usetikzlibrary{positioning}
\usetikzlibrary{shapes}
\usetikzlibrary{shapes.misc}
\usetikzlibrary{shapes.geometric}
\usetikzlibrary{plotmarks}
\usetikzlibrary{intersections}
\usetikzlibrary{calc}
\usetikzlibrary{fit}
\usetikzlibrary{patterns,tikzmark}
\usetikzlibrary{matrix,decorations.pathreplacing,calc}
\usepackage{amsthm}

\tikzset{cross/.style={cross out, draw, 
         minimum size=2*(#1-\pgflinewidth), 
         inner sep=0pt, outer sep=0pt}}

\newcommand{\ci}[0]{\mathbf{u}}

\newcommand{\x}[0]{\mathbf{x}}
\newcommand{\V}[0]{V}
\newcommand{\vt}[0]{\mathbf{v}_{t}}

\newcommand{\norm}[1]{\left\lVert#1\right\rVert}

\newcommand{\vertiii}[1]{{\left\vert\kern-0.25ex\left\vert\kern-0.25ex\left\vert #1 
    \right\vert\kern-0.25ex\right\vert\kern-0.25ex\right\vert}}

\newtheorem{theorem}{Theorem}
\newtheorem{proposition}{Proposition}

\newtheorem{corollary}{Corollary}[theorem]
\newtheorem{lemma}{Lemma}

\newtheorem{problem}{Problem}
\newtheorem{assumption}{Assumption}

\newtheorem{definition}{Definition}

\newmdenv[
    linecolor=white, backgroundcolor=lightgray!15, innertopmargin=5pt, innerbottommargin=5pt, skipabove=10pt, skipbelow=10pt
]{graybox}

\makeatletter
\renewcommand{\fps@figure}{htp}
\renewcommand{\fps@table}{htp}
\makeatother

\title{\LARGE \bf
Stochastic Neural Control Barrier Functions}

\author{ Hongchao Zhang$^{1}$, Manan Tayal$^{2}$, Jackson Cox$^{3}$, Pushpak Jagtap$^{4}$, Shishir Kolathaya$^{4}$ and Andrew Clark$^{3}$ %
\thanks{
This work was supported by National Science Foundation grant CMMI-2418806 and Air Force Office of Scientific Research grant FA9550-23-1-0208.
}
\thanks{$^{1}$ Department of Computer Science, Vanderbilt University, Nashville, TN 37235 USA
{\tt\scriptsize hongchao.zhang@vanderbilt.edu}.
}
\thanks{$^{2}$ Emergence AI.
{\tt\scriptsize mtayal@emergence.ai}
}
\thanks{$^{3}$ Electrical and Systems Engineering Department,  Washington University in St. Louis, St. Louis, MO 63130 USA
{\tt\scriptsize \{jackson.cox, andrewclark\}@wustl.edu}.
}%
\thanks{$^{4}$ Center for Cyber-Physical Systems, Indian Institute of Science (IISc), Bengaluru.
{\tt\scriptsize \{pushpak, shishirk\}@iisc.ac.in}.
}

}

\begin{document}

\maketitle

\begin{abstract}

Control Barrier Functions (CBFs) are utilized to ensure the safety of control systems. CBFs act as safety filters in order to provide safety guarantees without compromising system performance. These safety guarantees rely on the construction of valid CBFs. Due to their complexity, CBFs can be represented by neural networks, known as neural CBFs (NCBFs). Existing works on the verification of the NCBF focus on the synthesis and verification of NCBFs in deterministic settings, leaving the stochastic NCBFs (SNCBFs) less studied. In this work, we propose a verifiably safe synthesis for {SNCBFs with twice-differentiable (smooth) activation functions. We propose a verification-free synthesis framework and a verification-in-the-loop synthesis framework for smooth SNCBFs, and we validate our frameworks on two case studies, namely, the inverted pendulum and the unicycle model.}

\end{abstract}

\section{Introduction}
\label{sec:intro}

Safety is one of the fundamental properties required for control systems, especially those that interact with humans and critical infrastructures. Safety violations could lead to catastrophic damage to robots, harm to humans, and economic loss. The safety requirements of these systems, with applications including medicine, energy, and robotics, have motivated recent research to design safe control policies. Safety requirements can be formulated as the positive invariance of a given safe region, meaning the system remains in the safe region for all time \cite{ames2019control}.

Various approaches for safety-critical control have been proposed, including Hamilton-Jacobi Reachability (HJR) analysis \cite{tayal2025physics,choi2021robust} and safe reinforcement learning (RL) \cite{brunke2022safe,chow2019lyapunov}.
Scalability issues of HJR and the lack of guarantees of safe RL are open problems impeding the deployment of these methods on real-world applications. 
To combat these issues, control barrier function (CBF)-based approaches~\cite{ames2014control} have been utilized. These CBF-based approaches have been applied in deterministic ~\cite{ames2019control} and stochastic control systems \cite{jagtap2020formal,clark2021verification} due to their ease of implementation and compatibility with various safety and performance criteria \cite{dawson2023safe}. CBF-based quadratic programs (CBF-QPs) require known CBFs for the control system, which are not always given and need to be synthesized.

For control systems with polynomial dynamics, CBF synthesis can be formulated as sum-of-squares (SOS) constrained optimization problems \cite{clark2024semi,kang2023verification,dai2023convex}. However, SOS approaches require a polynomial barrier parameterization whose size scales combinatorially with state dimension and degree, becoming prohibitive for high-dimensional systems, and they do not apply to non-polynomial dynamics,
including learning-enabled robotic systems \cite{tampuu2020survey}. To address these limitations, Neural CBFs (NCBFs)~\cite{so2024train,dawson2022safe,dawson2023safe,liu2023safe,tayal2024learning} have been proposed to represent CBFs with feed-forward Neural Networks (NNs) by exploiting the uniform approximability of NNs.
NCBFs have shown substantial promise in applications including robotic manipulation \cite{dawson2022safe}, navigation \cite{xiao2023barriernet}, and flight control.

After synthesizing an NCBF, the NCBF must be verified to ensure that the resulting CBF-QP is feasible everywhere in the state space, and to ensure that the super-level-set of the NCBF is contained in the safe region~\cite{zhang2023exact,so2024train}.  Verification-In-The-Loop (VITL), also known as Counterexample-guided synthesis, has been implemented in neural barrier certificates~\cite{abate2021fossil,edwards2024fossil} and NCBF verification~\cite{wang2022design,zhang2025seev}. VITL synthesis of NCBFs is dependent on an effective and efficient verifier, which is an open problem due to scalability issues of verifying a neural network and its derivatives. 

Deterministic NCBFs are insufficient to ensure safety in the presence of random noise. Stochastic neural CBFs (SNCBFs) offer probabilistic forward invariance. However, the above challenge leaves their synthesis and verification an open problem. {SNCBFs were first studied in~\cite{zhang2024fault}, which stacked fault-tolerant SNCBFs across sensor-fault patterns, but the synthesized SNCBF still requires verification of both correctness and feasibility.}
To investigate this challenge, we propose a training framework to synthesize provably valid NCBFs for continuous-time, stochastic systems. Our methodology establishes a validity condition that ensures efficacy across the entire state space with only a finite number of data points. We train the network robustly by enforcing Lipschitz bounds on the neural network and its Jacobian and Hessian.

We make the following contributions towards synthesizing and verifying NCBFs for stochastic systems.

\begin{itemize}
    \item {We present a verification-free synthesis of smooth SNCBFs with a validity condition that certifies the SNCBF conditions over the entire state space from finitely many samples, with the required Lipschitz bounds on the network and its Jacobian and Hessian enforced via semidefinite programs.}
    \item {We develop a verification-in-the-loop (VITL) synthesis that generalizes the single-hidden-layer, verification-free approach to multiple hidden layers, formulating the verification of smooth SNCBFs with twice-differentiable activation functions as nonlinear programs solvable by Satisfiability Modulo Theories (SMT) solvers and relaxing the dense-sampling and single-layer assumptions.}
    \item {We validate our approach on the inverted pendulum and the unicycle model, showing that smooth SNCBFs output verifiably safe results while covering a larger safe subset than the baseline approach of a Fault-Tolerant SNCBF without VITL.}
\end{itemize}

The remainder of this paper is organized as follows. Section \ref{sec:preliminaries} presents the dynamics model and preliminary results. Section \ref{sec:SmoothSNCBF} presents the verifiable synthesis, verification, and VITL synthesis of SNCBFs with smooth activation functions. Section \ref{sec:experiments} presents the validation of the proposed methods with experimental settings, results, and comparison. Section \ref{sec:conclusion} concludes the paper.

\section{Preliminaries}
\label{sec:preliminaries}
In this section, we introduce the system model, stochastic control barrier functions (SCBFs), and needed mathematical background.
\begin{table}[h]
\vspace{-0.2em}
  \centering
    \begin{tabular}{@{}llll@{}}
        \toprule
        Symbol & Description & Symbol & Description \\
        \midrule
        $\mathcal{X}$    & State space          & $\alpha(\cdot)$ & Class $\mathcal{K}$ function \\
        $\mathcal{X}_S$  & Safe set             & $V$             & Diffusion coefficient \\
        $\mathcal{X}_U$  & Unsafe set           & $f,g$           & Drift, control vector fields \\
        $\mathcal{X}_I$  & Initial safe set     & $\theta$        & Neural-network parameters \\
        $\mathcal{D}$    & Super-0-level set of $B(\x)$ & $N$     & Sample size \\
        $B(\x)$          & CBF candidate        & $\bar{\epsilon}$ & Sampling radius \\
        $\mathcal{A}$    & Infinitesimal generator & & \\
        \bottomrule
    \end{tabular}
    \vspace{-0.2em}
    \caption{Notation Table}
\end{table}
\vspace{-2em}
\subsection{System Model}
\label{subsec:model}
We consider a continuous time stochastic control system with state $\x_{t} \in \mathcal{X}\subseteq\mathbb{R}^{n_{x}}$ and input $\ci_{t} \in \mathcal{U} \subseteq \mathbb{R}^{n_{u}}$ at time $t\geq0$ with a dynamic model formulated by the following stochastic differential equation (SDE): 
\begin{equation}
    \label{eq:state-sde}
    d\x_{t} = (f(\x_{t})+g(\x_{t})\ci_{t}) \ dt + \V \ d \vt,
\end{equation}
where $f: \mathbb{R}^{n_{x}} \rightarrow \mathbb{R}^{n_{x}}$ and $g: \mathbb{R}^{n_{x}} \rightarrow \mathbb{R}^{n_{x} \times n_{u}}$ are locally Lipschitz, $\vt$ is $n_{v}$-dimensional Brownian motion and $\V \in \mathbb{R}^{n_{x} \times n_{v}}$. 
We make the following assumptions on  system \eqref{eq:state-sde}. 
\begin{assumption}
    \label{assumption:sde_solution}
    There exists an initial state $\hat{\x}\in \mathcal{X}$ such that $\mathbb{P}[\x_{0}=\hat{\x}]=1$. The SDE \eqref{eq:state-sde} admits a unique strong solution. 
\end{assumption}
We denote the state space without obstacles as a safe set $\mathcal{X}_{S}$ and let $\mathcal{X}_{U} = \mathcal{X}/\mathcal{X}_{S}$ be the unsafe set. The safe set is the \textit{super-0-level set} of a differentiable function $h:\mathcal{X}\subseteq \mathbb{R}^{n_x} \rightarrow \mathbb{R}$, yielding
\begin{equation*}
	\mathcal{X}_{S} = \{ \mathbf{x}\in \mathcal{X} : h(\mathbf{x}) \geq 0\}, \quad  
	\mathcal{X}_{U} = \{ \mathbf{x}\in \mathcal{X} : h(\mathbf{x}) < 0\}.
\end{equation*}
We further let the interior and boundary of $\mathcal{X}_{S}$ be $\text{Int}\left(\mathcal{X}_{S} \right) = \{ \mathbf{x}\in \mathcal{X} : h(\mathbf{x}) > 0\}$ and $\partial\mathcal{X}_{S}  = \{ \mathbf{x}\in \mathcal{X} : h(\mathbf{x}) = 0\}$, respectively. 

\emph{Function Definitions: }We next present the definitions of class-$\kappa$ functions, extended class-$\kappa$ functions, and indicator functions. A continuous function $\alpha : [0, d) \rightarrow [0, \infty)$ for some $d > 0$ is said to belong to \emph{class-$\kappa$} if it is strictly increasing and $\alpha(0) = 0$. Here, $d$ is allowed to be $\infty$. If   function can be extended to the interval $\alpha: (-b,d)\to (-\infty, \infty)$ with $b>0$ (which is also allowed to be $\infty$),  we call it an \emph{extended class-$\kappa$ function} denoted as $\kappa_{\infty}$. We denote the \emph{indicator function} as $\mathds{1}_\mathcal{C}(\x)$ which takes the value of 1 if $\x\in \mathcal{C}$ and $0$ if $\x\notin\mathcal{C}$. Let $\omega$ denote a scalar variable and $a$ denote a constant scalar. We use the indicator function for logical operations, i.e., $\mathds{1}_{\omega > a}(x)$, which takes the value of 1 when $\omega > a$. Let $\left\{\omega-a\right\}_{+}:=\max \{\omega-a, 0\}$ and $\left\{\omega-a\right\}_{-}:=-\min \{\omega-a, 0\}$

\emph{Notations: } 
We denote the stochastic process of the state $\x$ as $\x_{t}$ and let $\x_{0}$ represent the initial state. Let $[\cdot]_{i}$ denote the $i$-th row of a matrix or the $i$-th item of a vector. For simplicity, we use $=\mathbf{0}$, and $\geq \mathbf{0}$ to represent the vector element-wise $=0$, and $\geq 0$, respectively.

\subsection{Preliminaries on Stochastic Processes}
We next present the infinitesimal generator, which we will use later to derive the safety conditions for our SNCBF.
\iffalse %
Any strong solution of an SDE is a semimartingale. Let $\omega_{t}$ denote a scalar continuous semimartingale process. The following Tanaka's formula explicitly links the local time of a semimartingale to its behavior.

\begin{theorem}[Tanaka's Formula \cite{revuz2013continuous}, Ch. 6, Theorem 1.2]
\label{def:TanakaFormula}
    For any real number $a$, there exists an increasing continuous process $L_t^a$ called the local time of $\omega_{t}$ in $a$, such that,
    $$
    \begin{aligned}
    \left|\omega_t-a\right| & =\left|\omega_0-a\right|+\int_0^t \operatorname{sgn}\left(\omega_s-a\right) d \omega_s+L_t^a \\
    \left\{\omega_t-a\right\}_{+} & =\left\{\omega_0-a\right\}_{+}+\int_0^t \mathds{1}_{\left(\omega_s>a\right)} d \omega+\frac{1}{2} L_t^a \\
    \left\{\omega_t-a\right)\}_{-} & =\left\{\omega_0-a\right\}_{-}-\int_0^t \mathds{1}_{\left(\omega_s \leq a\right)} d \omega_s+\frac{1}{2} L_t^a.
    \end{aligned}
    $$
\end{theorem}
A detailed treatment of  semimartingales and Tanaka's formula can be found in \cite{revuz2013continuous}.
\fi %
Consider a  function $B: \mathbb{R}^{n_{x}} \rightarrow \mathbb{R}$ mapping from a state $\x$ to a scalar.
The infinitesimal generator of $B$ is defined as follows. 
\begin{definition}[Infinitesimal generator]
Let $\x_{t}$ be the strong solution to \eqref{eq:state-sde}. The infinitesimal generator $\mathcal{A}$ of $\x_t$ is defined by
$$
\mathcal{A} B(\x)=\lim _{t \downarrow 0} \frac{\mathbb{E}\left[B\left(\x_t\right)\mid \x_{0}=\x\right]-B(\x)}{t} ; \hspace{.25cm}\x \in \mathbb{R}^{n_{x}}
$$
where $B(\x)$ is in a set of functions $\mathcal{D}(\mathcal{A})$ (called the domain of the operator $\mathcal{A}$) such that the limit exists at $\x$.
\end{definition}

We make the following assumption on $B(\x)$. 
\begin{assumption}
    \label{assumption:domain_operator}
    We assume the function $B(\x)$ is in the domain of the operator $\mathcal{D}(\mathcal{A})$. Specifically, the following two conditions are satisfied in $\mathcal{D}(\mathcal{A})$. 
    \begin{itemize}
        \item The expectation $\mathbb{E}\left[B\left(\x_{t}\right)\mid \x_{0}=\x\right]$ is well-defined and finite. 
        \item The limit defining $\mathcal{A} B(\x)$ exists for every $\x \in \mathcal{X}$.
    \end{itemize}
\end{assumption}
We note that $\mathcal{D}(\mathcal{A})=\mathbb{R}^{n_{x}}$ if $B(\x)$ is twice differentiable. 
\subsection{Stochastic Control Barrier Functions}
The control policy must ensure that the system adheres to a safety constraint defined by the positive invariance of a specified safety region $\mathcal{X}_{S}$.

\subsubsection{Control Barrier Functions for Stochastic Systems}

We introduce the stochastic CBF (SCBF) for stochastic systems by presenting the definition of stochastic (zeroing) CBF and SCBF-QP. 
\begin{definition}[Stochastic Zeroing CBF]
    \label{def:SZCBF}
    A continuously differentiable function $B: \mathcal{X} \rightarrow \mathbb{R}$ is called a stochastic zeroing control barrier function for system \eqref{eq:state-sde} if the following conditions are satisfied. 
    \begin{itemize}
        \item $B(\x) \in \mathcal{D}(\mathcal{A})$;
        \item $B(\x) \geq 0$ for all $\x \in \mathcal{D} \subseteq \mathcal{X}_{S}$;
        \item $B(\x)<0$ for all $\x \notin \mathcal{D}$;
        \item there exists an extended $\kappa_{\infty}$ function $\alpha$ such that
        $$
        \sup _{u \in \mathcal{U}}[\mathcal{A} B(\x)+\alpha(B(\x))] \geq 0.
        $$
    \end{itemize}
\end{definition}
Let $\mathcal{D}:=\{\x:B(\x)\geq 0\}$ to denote the super-0-level-set of the function $B(\x)$.  
The SCBF-QPs are regarded as \textit{safety filters} which take the control input from a reference controller $\ci_{ref}(\x,t)$ and modify it so that it abides by our constraint of positive invariance:
\begin{equation}
\begin{aligned}
\label{eqn: CBF QP}
\ci^{*}(\x,t) &= \min_{\ci \in \mathcal{U}} \norm{\ci - \ci_{ref}(\x,t)}^2\\
\quad & \textrm{s.t. } \mathcal{A} B(\x)+\alpha(B(\x)) \geq 0.
\end{aligned}
\end{equation}
Control signals satisfying SCBF-QP ensure probabilistic positive invariance of the set $\mathcal{D}$. 

Next, we present the worst-case probability estimation. 
\begin{proposition}
[{\cite[Proposition III.5]{wang2021safety}}]
\label{prop:safety-finite-time}
Suppose the map $B(\x)$ is a Stochastic CBF with a linear class-$\alpha$ function (where $\alpha(\x) = k\x; k>0$), and the control strategy as $\mathcal{U}_{z}=\{u \in \mathcal{U}: \mathcal{A} B(\x)+k B(\x) \geq 0\}$. Let $c=\sup _{x \in \mathcal{C}} B(\x)$ and $\x_{0} \in int(\mathcal{D})$, then under any $\ci \in \mathcal{U}_{z}$ we have the following worst-case probability estimation:
$$
\mathbb{P}\left[\x_t \in int(\mathcal{D}), 0 \leq t \leq T\right] \geq\left(\frac{B(\x_{0})}{c}\right) e^{-c T}.
$$
\end{proposition}

\subsubsection{Neural CBFs}
\label{subsubsec:NCBFs}

CBFs defined by neural networks, denoted Neural CBFs (NCBFs), leverage the expressiveness of NNs; we let $\theta$ denote the network parameters. An NCBF synthesized for a stochastic system is referred to as a stochastic neural CBF (SNCBF).

%

\subsection{Preliminary Results}
We next present preliminary results from Positivestellensatz and Farkas' Lemma. These results will be used to derive the verification conditions for SNCBFs.
The polynomial $p(\x)$ is Sum-Of-Squares (SOS) if and only if it can be written as 
$p(\x) = \sum_{i=1}^{n} (p_i(\x))^2$,
for some polynomials $p_i(\mathbf{x})$. 

We next present the required background on real algebraic geometry. 
For certifying non-negative polynomials, the cone is given by $\Sigma[\phi_{1},\ldots,\phi_{k_{\Sigma}}] =$
\begin{equation*}
\left\{\sum_{K \subseteq \{1,\ldots,k_{\Sigma}\}}{\gamma_{K}(\mathbf{x})\prod_{i \in K}{\phi_{i}(\mathbf{x})}} : \gamma_{K}(\mathbf{x}) \in \mbox{ SOS } \forall K \subseteq \{1,\ldots,k_{\Sigma}\}\right\}.
\end{equation*}
The monoid $\mathcal{M}$ generated from a set of polynomials $\chi_{1},\ldots,\chi_{k_{\mathcal{M}}}$ with non-negative integer exponents $r_{1}\ldots r_{k_{\mathcal{M}}}$ is defined as
\begin{equation*}
\mathcal{M}[\chi_{1},\ldots,\chi_{k_{\mathcal{M}}}] = \left\{{\prod_{i =1}^{k_{\mathcal{M}}}{\chi^{r_{i}}(\mathbf{x})}} : r_{1}\ldots r_{k_{\mathcal{M}}}\in\mathbb{Z}_{+}\right\}.
\end{equation*}
The ideal generated from polynomials $\Gamma_{1},\ldots,\Gamma_{k_{\mathbb{I}}}$ is given by 
\begin{equation*}
\mathbb{I}[\Gamma_{1},\ldots,\Gamma_{k_{\mathbb{I}}}]
= \left\{\sum_{i=1}^{k_{\mathbb{I}}}{p_{i}(\mathbf{x})\Gamma_{i}(\mathbf{x})} : p_{1},\ldots,p_{k_{\mathbb{I}}} \mbox{ are polynomials}\right\}.
\end{equation*}

The following theorem shows that the emptiness of a set formed by the polynomials described above is equivalent to the existence of polynomials within their respective cone, square of monoid, and ideal such that their sum is equal to zero. 
\begin{theorem}[Positivstellensatz \cite{parrilo2003semidefinite}]
\label{th:Positivstellensatz}
Let $\left(\phi_{i}\right)_{j=1, \ldots, k_{\Sigma}}$, $\left(\chi_{j}\right)_{k=1, \ldots, k_{\mathcal{M}}}$, $\left(\Gamma_{\ell}\right)_{\ell=1, \ldots, k_{\mathbb{I}}}$ be finite families of polynomials. Then, the following properties are equivalent:
\begin{enumerate}
    \item The following set is empty.
    \begin{equation}
    \label{eq:psatz_set}
    \left\{
    \begin{array}{l|ll}
    \x \in \mathbb{R}^{n_{x}} &
    \begin{array}{ll}
    \phi_{i}(\x) \geq 0, & i=1, \ldots, k_{\Sigma} \\
    \chi_{j}(\x) \neq 0, & j=1, \ldots, k_{\mathcal{M}} \\
    \Gamma_{l}(\x)=0, & l=1, \ldots, k_{\mathbb{I}}
    \end{array}
    \end{array}
    \right\}
    \end{equation}
    \item There exist $\phi \in \Sigma, \chi \in \mathcal{M}, \Gamma \in \mathbb{I}$ such that $\phi+\chi^{2}+\Gamma=0$.
\end{enumerate}
\end{theorem}

The following  is one of the variants of Farkas' Lemma. 
\begin{lemma}[Farkas' Lemma {\cite[Proposition 6.4.3)]{matousek2006understanding}}]
\label{Lemma:Farkas}
Let $A$ be a real matrix with $m$ rows and $n$ columns, and let $b \in \mathbb{R}^n$ be a vector. Exactly one of the following conditions holds (the conditions are mutually exclusive):
\begin{itemize}
    \item The system of inequalities $A {x} \leq {b}$ has a solution
    \item There exists ${y}$ such that ${y} \geq \boldsymbol{0}$, ${y}^{T} A=\boldsymbol{0}^{T}$ and ${y}^{T} {b} < 0$, where $\boldsymbol{0}$ denote a zero vector.
\end{itemize}
\end{lemma}

\section{Smooth Stochastic Neural CBF}
\label{sec:SmoothSNCBF}

In this section, we present the synthesis of SNCBF with smooth activation functions.  Our approach assumes that the NN employs smooth (i.e., twice differentiable) activation functions so that its Jacobian and Hessian are well-defined. The synthesis procedure is developed in two frameworks: (i) verifiable synthesis and (ii) VITL synthesis with a novel verification for smooth SNCBFs.

\begin{problem}
\label{prob:smooth_sncbf}
    Given a continuous-time stochastic control system defined as \eqref{eq:state-sde}, a desired confidence level $\eta$, state space $\mathcal{X}$, initial safe set and the unsafe set $\mathcal{X}_{I}\subseteq\mathcal{X}_{S}$ and $\mathcal{X}_{U}=\mathcal{X}\setminus\mathcal{X}_S$, respectively, devise an algorithm that synthesizes a stochastic neural CBF $(\mathrm{SNCBF})$ $B(\x)$ with continuously differentiable (smooth) activation functions. In particular, the SNCBF must satisfy the following worst-case probability guarantee:
$$
\mathbb{P}\left[\x_t \in \mathcal{X}_S, \forall t\in [0, T] \mid \x_{0}\in int(\mathcal{D})\right] \geq 1- \eta.$$
\end{problem}

Our strategy to address Problem \ref{prob:smooth_sncbf} is to (i) perform a verifiable synthesis of the SNCBF under a set of initial assumptions, (ii) relax some of these assumptions by introducing verification conditions, and (iii) propose a synthesis loss function that embeds verification into the learning loop. 

To facilitate the synthesis and verification of an SNCBF that meets the safety criteria in Problem~\ref{prob:smooth_sncbf}, it is essential to analyze how the barrier function evolves along the trajectories of the stochastic system. In continuous-time settings, the following lemma provides the precise expression for the infinitesimal generator of an SNCBF with smooth activation functions:

The infinitesimal generator of an SNCBF $(\mathrm{SNCBF})$ $B(\x)$ with twice-differentiable  activation functions is given by \cite{so2023almost} as: 
\begin{equation}
    \label{eq:Ah_tilde}
    \mathcal{A}B= \frac{\partial B}{\partial\x}\left(f(\x)+g(\x)\ci\right) + 
    \frac{1}{2}\mathsf{tr}\left(\V^\intercal \frac{\partial^2 B}{\partial \x^2} \V\right).
\end{equation}

\subsection{Smooth SNCBF Verifiable Synthesis}
\label{subsec:verifiable_synth}

The conditions of Problem \ref{prob:smooth_sncbf} are satisfied if we can synthesize a NN $B(\x)$ parameterized by $\theta$ to represent our SCBF.
We denote control policy $\mu: \mathbb{R}^{n_x} \rightarrow \mathbb{R}^{n_u}$ as a mapping from the  state $\x_{t}$ to a control input $\ci_{t}=\mu(\x_{t})$ at each time $t$. $B(\x)$ ensures safety by imposing conditions on system states and control inputs defined as follows. 
\begin{align}
    & B(\x) \geq 0, \forall \x\in \mathcal{X}_{I}, \notag\\
    & B(\x) < 0,  \forall \x\in \mathcal{X}_{U}, \notag\\
    & \mathcal{A}B(\x) +  \alpha\left(B(\x)\right) \geq 0,  \forall \x\in \mathcal{D}.\label{eq: problem1}
\end{align}

To approach Problem \ref{prob:smooth_sncbf}, we make the following assumptions in Section \ref{subsec:verifiable_synth}.  
\begin{assumption}
\label{Assumption:SmoothSNCBF} 
For the smooth SNCBF verifiable synthesis, we assume that
\begin{enumerate}
    \item The function $B(\x)$, $\frac{\partial B}{\partial \x}$, $\frac{\partial^{2} B}{\partial \x^{2}}$ and control policy $\mu$ are locally Lipschitz continuous.
    \item The SNCBF $B(\x)$ is a feedforward NN with one hidden layer and twice differentiable activation function. 
    \item $\V$ is a $n_{x}\times n_{x}$ diagonal matrix. 
    \item The derivative of each activation function is lower bounded by $\underline{\sigma^{\prime}}$ and upper bounded by $\overline{\sigma^{\prime}}$. 
    \end{enumerate}
\end{assumption}

Next, we formulate our verifiable synthesis of $B(\x)$ as a robust optimization problem (ROP) with auxiliary variable $\psi$:

\begin{equation}\label{eq: rcp}
    \mathrm{ROP}: \begin{cases}\underset{\psi}{\min } & \psi \\ \text { s.t. } & \max \left(q_{k}(\x)\right) \leq \psi, k \in\{1,2,3\} \\ & \forall \x \in \mathcal{X}, \psi \in \mathbb{R},\end{cases}
\end{equation}

where
\begin{equation} \label{eq:q_conditions}
\begin{aligned}
    q_{1}(\x)=& \left(-B(\x)\right) \mathds{1}_{\mathcal{X}_{I}}(\x), \\
    q_{2}(\x)=& \left(B(\x)+\delta \right) \mathds{1}_{\mathcal{X}_{U}}(\x), \\
    q_{3}(\x)=& -\mathcal{A}B(\x)- \alpha(B(\x)),
\end{aligned}
\end{equation}
where $\delta$ is a small positive scalar ensuring strict inequality. The ROP in \eqref{eq: rcp} inherently involves infinitely many constraints due to the continuous nature of the state space $\mathcal{X}$.

To approximate the solution, we employ the scenario optimization program (SOP), which replaces the ROP with a finite number of sampled constraints. Given a scalar $\bar{\epsilon}$, we uniformly sample $N$ data points $\x_{i} \in \mathcal{X}, i \in\{1, \ldots, N\}$. 
Let $\bar{\epsilon}$ be a positive scalar. We sample points dense enough such that $\left\|\x-\x_{i}\right\| \leq \bar{\epsilon}$ for any $\x\in\mathcal{D}$ and some $\x_{i}$. We then solve:
\begin{equation}\label{eq: scp}
    \mathrm{SOP}: \begin{cases}\underset{\psi}{\min } & \psi \\ \text { s.t. } & q_{1}\left(\x_{i}\right) \leq \psi, \forall \x_{i} \in \mathcal{S}, \\ & q_{2}\left(\x_{i}\right) \leq \psi, \forall \x_{i} \in \mathcal{U}, \\ & q_{3}\left(\x_{i}\right) \leq \psi, \forall \x_{i} \in \mathcal{D}, \\ & \psi \in \mathbb{R}\end{cases}
\end{equation}
where $q_{k}(\x), k \in{1,2,3}$ are as defined in \eqref{eq:q_conditions}, and the data sets $\mathcal{S}, \mathcal{U}$, and $\mathcal{D}$ correspond to points sampled from the initial safe set $\mathcal{X}_{I}$, initial unsafe set $\mathcal{X}_{U}$, and state space $\mathcal{X}$, respectively.

Since SOP is a linear program in the decision variable $\psi$, a feasible solution can be obtained, denoted as $\psi^{*}$. The following theorem establishes conditions ensuring that the SNCBF $B(\x)$ obtained via \eqref{eq: scp} provides a valid solution to Problem \ref{prob:smooth_sncbf}.

\begin{theorem}\label{thm: smooth_ncbf}

Consider a continuous time stochastic control system \eqref{eq:state-sde}, and initial safe and unsafe sets $\mathcal{X}_{I} \subseteq\mathcal{X}_{S} \subseteq \mathcal{X}$ and $\mathcal{X}_{U} \subseteq \mathcal{X}\setminus\mathcal{X}_S$, respectively. Let  $B(\x)$ be the SNCBF with trainable parameters $\theta$. 
Suppose condition 1) in Assumption \ref{Assumption:SmoothSNCBF} holds so that the functions $q_{k}(\x), k \in\{1,2,3\}$ in equation \eqref{eq:q_conditions} are Lipschitz continuous. 
Given $\bar{\epsilon}$, suppose we sample $N$ data points $\x_1, \ldots, \x_N \in \mathcal{X}$ such that for any point $\x$ in the state space, there exists a sampled point $\x_i$ satisfying $\|\x - \x_i\| \leq \bar{\epsilon}$.
For the SOP \eqref{eq: scp} constructed by utilizing $\x_1,\ldots,\x_N$ such that $\mathcal{D}$ is covered by balls centered at the $\x_i$'s with radii $\bar{\epsilon}$, let $\psi^{*}$ be the optimal value. Then $B(\x)$ is a valid $\mathrm{SNCBF}$, i.e., it solves Problem~\ref{prob:smooth_sncbf}, if the following condition holds:
\begin{equation} \label{eq: completeness_condition}
    L_{\max} \bar{\epsilon}+\psi^{*} \leq 0,
\end{equation}
where $L_{\max}$ is the maximum of the Lipschitz constants of $q_{k}(\x), k \in\{1,2,3\}$ in \eqref{eq:q_conditions}.
\end{theorem}
\begin{proof}
For any $\x$ and any $k\in\{1,2,3\}$, we know that:
\begin{equation*} 
    \begin{aligned}
    q_k(\x) & =  q_k(\x) - q_k(\x_{i}) + q_k(\x_{i})\\
    & \leq L_{k}\left\| \x - \x_{i} \right\| + \psi^*\\
    & \leq L_{k}\bar{\epsilon} + \psi^* \leq L_{\max}\bar{\epsilon} + \psi^* \leq 0.
    \end{aligned}
\end{equation*}
 Hence, if $q_{k}(\x), k \in\{1,2,3\}$ satisfies condition \eqref{eq: completeness_condition}, then the $B(\x)$ is a valid $\mathrm{SCBF}$, satisfying conditions \eqref{eq: problem1}. 
\end{proof}

Theorem \ref{thm: smooth_ncbf} reformulates Problem \ref{prob:smooth_sncbf} as follows. Given the data sets $\mathcal{S}, \mathcal{U}$ and $\mathcal{D}$, devise an algorithm that synthesizes SNCBF $B(x)$ such that it satisfies the conditions required in SOP \eqref{eq: scp} and $\psi^*$ satisfies condition \eqref{eq: completeness_condition}.

We next present the method to synthesize a smooth SNCBF. %
Let $\mathbf{e}_{i}$ denote a one-hot vector, e.g., $[\mathbf{e}_{i}]_{i} = 1$ and $[\mathbf{e}_{i}]_{j} = 0$ for $j \neq i$.  We define a diagonal weighting matrix as follows. 
$$\Omega :=\left\{\Omega \in \mathbb{R}^{n \times n} \mid \Omega=\sum_{i=1}^n \omega_{i i} \mathbf{e}_{i} \mathbf{e}_{i}^T, \omega_{ii} \geq 0\right\}.$$

For an NN with one hidden layer, we have $\Omega  \in \mathcal{D}_{n_{x}}$. By \cite[Section III]{pauli2021training}, the NN is Lipschitz continuous with coefficient $L$ if there exists a nonnegative diagonal matrix $\Omega$ such that  the matrix $M(\theta,\Omega)$ defined by  
\begin{equation*}
    M(\cdot) =
    \begin{bmatrix} 
    L^2 \textbf{I} + 2 \underline{\sigma^{\prime}}\overline{\sigma^{\prime}}\mathbf{W}_{1}^T\Omega\mathbf{W}_{1} & -(\underline{\sigma^{\prime}} + \overline{\sigma^{\prime}})\mathbf{W}_{1}^T\Omega  & 0\\
    -(\underline{\sigma^{\prime}} + \overline{\sigma^{\prime}})\Omega\mathbf{W}_{1} & 2\Omega & -\mathbf{W}_{2}\\
    0 & -\mathbf{W}_{2} &  \textbf{I}
    \end{bmatrix}
\end{equation*}
is positive semidefinite.

Our scenario necessitates ensuring not only the Lipschitz boundedness of the SNCBF $B(\x)$, but also of $\frac{\partial B}{\partial \x}$ and $\V^\intercal \frac{\partial^2 B}{\partial \x^2} \V$. Therefore, we must explore the relationship between the network weights and the matrix $M$ to guarantee the boundedness of the aforementioned terms. To address this issue, we introduce the following theorem
which presents a method to ensure the L-Lipschitz continuity of the SNCBF $B(\x)$, its derivative, and the Hessian terms in \eqref{eq:q_conditions}.
\begin{theorem} 
\label{thm: effective_weights}
    Suppose Assumption \ref{Assumption:SmoothSNCBF} holds for an SNCBF $B(\x)$. The certificate for  L-Lipschitz continuity of $\frac{\partial B}{\partial \x}$ which is the derivative of the NN is given by $M_{\hat\sigma}(\hat\theta, \Omega)\succeq0$, where $\hat\sigma= \sigma'$ and  $\hat\theta = (\mathbf{W}_{1}, \hat{\mathbf{W}}_{2})$, $\hat{\mathbf{W}}_{2}$ is defined as:
    \begin{equation}
        \hat{\mathbf{W}}_{2} = \mathbf{W}_{1}^{T} \mathsf{diag}(\mathbf{W}_{2}).
    \end{equation}
    
    Additionally, the certificate for the L-Lipschitz continuity of $\mathsf{tr}(\V^T\frac{\partial^2 \mathbf{y}}{\partial x^2}\V)$ is expressed as $M_{\bar\sigma}(\bar\theta, \Omega)\succeq0$, where $\bar\sigma= \sigma''$ and $\bar\theta = (\mathbf{W}_{1},\bar{\mathbf{W}}_{2})$ and $\bar{\mathbf{W}}_{2}$ is defined as:
    \begin{equation}
        \bar{\mathbf{W}}_{2} = \begin{bmatrix} \sum_{j=0}^{r}\V_j^2\mathbf{W}_{2}^{j1}\mathbf{W}_{1}^{1j} & \dots & \sum_{j=0}^{r}\V_j^2\mathbf{W}_{2}^{jn}\mathbf{W}_{1}^{pj}
    \end{bmatrix}.
    \end{equation}
\end{theorem}
\begin{proof}
    Consider an NN with $p$ neurons in a hidden layer. The dimension of the input ($\x$) is $n_{x} \times 1$, the dimension of pre-final weight ($\mathbf{W}_{1}$) is $p \times n_{x} $, the dimension of pre-final bias ($\mathbf{r}_{1}$) is $p \times 1$ and the dimension of final weight ($\mathbf{W}_{2}$) is $1 \times p$. Let us start by  differentiating the NN: 
    \begin{equation*}
    \begin{aligned}
        \mathbf{y} &= \mathbf{W}_{2} \sigma(\mathbf{W}_{1} x + \mathbf{r}_{1}) \\
        \frac{\partial \mathbf{y}}{\partial \x} &= \mathbf{W}_{2} \mathsf{diag}(\sigma')\mathbf{W}_{1} \ (\text{Here, } \sigma' = \sigma'(\mathbf{W}_{1} \x + \mathbf{r}_{1})).\\
        \end{aligned}
    \end{equation*}
    The dimension of $\frac{\partial \mathbf{y}}{\partial \x}$ is $1 \times n_{x}$, therefore, its transpose will have the dimension of $n_{x} \times 1$. 
    \begin{equation*}
    \begin{aligned}
        (\frac{\partial \mathbf{y}}{\partial \x})^T = (\mathbf{W}_{2} \mathsf{diag}(\sigma')\mathbf{W}_{1})^T 
        =\underbrace{\mathbf{W}_{1}^{T} \mathsf{diag}(\mathbf{W}_{2})}_{\hat\theta_l} \sigma'(\mathbf{W}_{1} x + \mathbf{r}_{1}).\\
        \end{aligned}
    \end{equation*}
The term derivative term ($\frac{\partial \mathbf{y}}{\partial \x}$) is equivalent to an NN with with activation $\hat\sigma=  \sigma_l'$ and weight parameters $\hat\theta = (\mathbf{W}_{1},\hat{\mathbf{W}}_{2})$.

Therefore, the certificate for  L-Lipschitz continuity of the derivative term ($\frac{\partial \mathbf{y}}{\partial \x}$) is given by $M_{\hat\sigma}(\hat\theta, \Omega)\succeq0$.
Now, let us  calculate $\mathsf{tr}(\V^T\frac{\partial^2 \mathbf{y}}{\partial \x^2}\V)$, where $\V$ is $n_{x} \times n_{x}$ diagonal matrix.

\begin{equation*}
    \begin{aligned}
        \frac{\partial \mathbf{y}}{\partial \x} &= \hat{\mathbf{W}}_{2} \hat\sigma(\mathbf{W}_{1} x + \mathbf{r}_{1}) \\
        \V^T\frac{\partial^2 \mathbf{y}}{\partial \x^2}\V &= \V^T(\hat{\mathbf{W}}_{2} \mathsf{diag}(\hat\sigma')\mathbf{W}_{1} )\V\\
        \mathsf{tr}(\V^T\frac{\partial^2 \mathbf{y}}{\partial \x^2}\V) &= \mathsf{tr}(\V^T(\hat{\mathbf{W}}_{2} \mathsf{diag}(\hat\sigma')\mathbf{W}_{1} )\V)\\
    \end{aligned}
\end{equation*}

\begin{equation*}
    =\underbrace{\begin{bmatrix} \sum_{j=0}^{r}\V_j^2\hat{\mathbf{W}}_{2}^{j1}\mathbf{W}_{1}^{1j} \ldots \sum_{j=0}^{r}\V_j^2\hat{\mathbf{W}}_{2}^{jp}\mathbf{W}_{1}^{pj}
    \end{bmatrix}}_{\bar{\mathbf{W}}_{2}} \sigma^{\prime\prime}(\mathbf{W}_{1} \x + \mathbf{r}_{1}).
\end{equation*}

The term $\mathsf{tr}(\V^T\frac{\partial^2 \mathbf{y}}{\partial \x^2}\V)$ is equivalent to a feedforward NN with activation $\bar\sigma= \sigma''$ and weight parameters $\bar\theta = (\mathbf{W}_{1},\bar{\mathbf{W}}_{2})$ and $\hat{\mathbf{W}}_{2}$ is defined as:
    \begin{equation*}
        \bar{\mathbf{W}}_{2} = \begin{bmatrix} \sum_{j=0}^{n_{x}}\V_j^2\hat{\mathbf{W}}_{2}^{j1}\mathbf{W}_{1}^{1j} & \dots & \sum_{j=0}^{n_{x}}\V_j^2\hat{\mathbf{W}}_{2}^{jp}\mathbf{W}_{1}^{pj}
    \end{bmatrix}.
    \end{equation*}

Therefore, the certificate for Lipschitz continuity of the $\mathsf{tr}(\V^T\frac{\partial^2 \mathbf{y}}{\partial \x^2}\V)$ term is given by $M_{\bar\sigma}(\bar\theta, \Omega)\succeq0$.
\end{proof}

To train the SNCBF to satisfy (10), we define a loss function with three terms matching the three constraints. The first penalizes $B(\x_i) < 0$ for $\x_i \in \mathcal{X}_I \subseteq \mathcal{X}_S$, ensuring non-negativity in the initial safe set. The second penalizes $B(\x_i) \geq -\delta$ for $\x_i \in \mathcal{X}_U$, ensuring the barrier is sufficiently negative in the unsafe set. The third introduces a slack variable $\psi$, allowing for small constraint violations, quantifies infeasibility.
\begin{equation}
\label{eq:loss_cons}
    \begin{aligned}
        \mathcal{L}(\theta) = &\frac{1}{N} \sum_{\x_{i} \in \mathcal{S}} \max \left(0,\left(-B(\x_{i})\right) \mathds{1}_{\mathcal{X}_{I}} - \psi \right), \\
        + &\frac{1}{N} \sum_{\x_{i} \in \mathcal{U}} \max \left(0, \left(B(\x_{i})+\delta \right) \mathds{1}_{\mathcal{X}_{U}} - \psi\right), \\
        + &\frac{1}{N} \sum_{\x_{i} \in \mathcal{D}} \max (0, -\frac{\partial B}{\partial \x} f(\x_{i}) + \frac{\partial B}{\partial \x}g(\x_{i})\ci + \\
    &\frac{1}{2}\mathsf{tr}\left(\V^\intercal \frac{\partial^2 B(\x_{i})}{\partial \x^2} \V\right) -  \alpha(B(\x_{i})) - \psi ).
    \end{aligned}
\end{equation}

Let us consider a constrained optimization problem aiming to minimize loss $\mathcal{L}(\theta)$ in \eqref{eq:loss_cons} subject to $M_{j}(\theta, \Omega) \succeq 0$ for $j=1, 2, 3$, to ensure the Lipschitz continuity of the SNCBF $B(\x)$, and its derivative and Hessian terms. By employing a log-determinant barrier function, we convert this into an unconstrained optimization problem:
\begin{equation*}
\min _{\theta, \Omega} \mathcal{L}\left(f_{\theta}\right) + \mathcal{L}_{M}(\theta, \Omega),
\end{equation*}
where $\mathcal{L}_{M}(\theta, \Omega) = -\sum_{j=0}^{q} \rho_{j} \log \operatorname{det}\left(M_{j}(\theta, \Omega)\right)$ and $\rho_{j}>0$ are barrier parameters. Ensuring that the loss function $\mathcal{L}_{M}(\theta, \Omega) \leq 0$, guarantees that the linear matrix inequalities $M_{j}(\theta, \Omega) \succeq 0, j=1, 2, 3$ hold true. 
Let us consider the loss functions characterizing the satisfaction of Lipschitz bound as
\begin{equation}\label{eq: lmi_loss}
    \begin{aligned}
     \mathcal{L}_{M}&(\theta, \Omega, \hat\Omega, \bar\Omega)=-c_{l_{1}} \log \operatorname{det}(M_1(\theta, \Omega)) \\ &-c_{l_{2}} \log \operatorname{det}(M_2(\hat\theta, \hat\Omega))
      -c_{l_{3}} \log \operatorname{det}(M_3(\bar\theta, \bar\Omega)),
    \end{aligned}
\end{equation}
where $c_{l_{1}}, c_{l_{2}}, c_{l_{3}}$ are positive weight coefficients for the sub-loss functions, $M_1, M_2, M_3$ are the semi-definite matrices corresponding to the Lipschitz bounds $L_{h}, L_{dh}, L_{d2h}$ respectively, $\Omega, \hat\Omega, \bar\Omega$ are trainable parameters and $\theta, \hat\theta, \bar\theta$ are the weights mentioned in Theorem \ref{thm: effective_weights}.

Finally, let us consider the following loss function to satisfy validity condition \eqref{eq: completeness_condition}:
\begin{equation}\label{eq: completeness_loss}
    \begin{aligned}
    \mathcal{L}_{v}(\psi)=\max \left(0,L_{\max } \bar{\epsilon}+\psi\right),
    \end{aligned}
\end{equation}
where $L_{\max}$ is maximum of the Lipschitz constants of $q_{k}(x), k \in\{1,2,3\}$ in \eqref{eq:q_conditions}, or $L_{\max}=\max \left(L_{h}, L_{h}+L_{dh}L_{x} + L_{d2h}\right)$ and $\bar{\epsilon}$ is our sampling density. 

The training procedure begins by fixing all hyperparameters, including $\bar{\epsilon}, \mathcal{L}_{h}, \mathcal{L}_{dh}, \mathcal{L}_{d2h}, \omega_{1}, \omega_{2}, c_{l_{1}}, c_{l_{2}}, c_{l_{3}}$, and the maximum number of epochs. The $\texttt{Learn}(\mathcal{L}, \cdot)$ function performs gradient-based optimization to minimize loss $\mathcal{L}$ with respect to its second argument (network parameters, Lipschitz bound parameters, or slack variables). We employ stochastic gradient descent (SGD) with learning rate $\lambda$ for $K$ iterations.
The overall algorithm is summarized in Algorithm \ref{alg:sncbf}.
\vspace{-0.5em}
\begin{algorithm}
\caption{Training Formally Verified SNCBF}\label{alg:sncbf}
\begin{algorithmic}
\Require Data Sets: $\mathcal{S}, \mathcal{U}, \mathcal{D}$, Dynamics: $f, g, \sigma$, Lipschitz Bounds: $L_h, L_{dh}, L_x, L_{d2h}$
\State Initialize($\theta, \psi, \Omega, \Omega', \Omega''$)
\State $x_i \gets sample(\mathcal{S}, \mathcal{U},\mathcal{D})$
\State $L_{\max} \gets L_h, L_{dh}, L_x, L_{d2h}$
\While{$\mathcal{L}_{\theta} > 0$ or $\mathcal{L}_M \not\leq 0$ or $\mathcal{L}_v > 0 $ }
    \State $B \gets \theta$
    \State $u_i \gets \mathrm{SNCBF\_QP}(B, f, g, \sigma, \x_i, \psi)$ 
    \Comment{From eq. \hspace{.18cm}\eqref{eqn: CBF QP}}
    \State $\mathcal{L}_{\theta} \gets (B, f, g, \sigma, \x_i, \ci_i, \psi)$ \Comment{From eq. \eqref{eq:loss_cons}}
    \State $\theta \gets Learn(\mathcal{L}_{\theta}, \theta)$
    \State $\mathcal{L}_{M} \gets (\theta, \Omega, \Omega', \Omega'')$ \Comment{From eq. \eqref{eq: lmi_loss}}
    \State $\theta, \Omega, \Omega', \Omega'' \gets Learn(\mathcal{L}_{M})$
    \State $\mathcal{L}_{v} \gets (L_{\max}, \psi)$ \Comment{From eq. \eqref{eq: completeness_loss}}
    \State $\psi \gets Learn(\mathcal{L}_{v})$
\EndWhile
\end{algorithmic}
\end{algorithm}
\vspace{-0.5em}
We note that the verification-free synthesis approach requires one hidden layer (Assumption \ref{Assumption:SmoothSNCBF}) and dense sampled training dataset. Section \ref{subsec:smooth_SNCBF_veri} relaxed both assumptions by introducing VITL synthesis with SMT-based verification, which relaxes the dense sampling requirement.

\subsection{Smooth SNCBF Verification and Synthesis}
\label{subsec:smooth_SNCBF_veri}

The previous approach is limited to single hidden layer NNs and requires dense sampling under Assumption \ref{Assumption:SmoothSNCBF}. 
In this subsection, we consider an NN with multiple hidden layers. To handle the increased complexity, we employ a verification-in-the-loop (VITL) framework that iteratively refines the SNCBF parameters based on verification results. This approach relaxes Assumption \ref{Assumption:SmoothSNCBF}, allowing for more expressive NN architectures while still ensuring safety.
The safety guarantee of our smooth SNCBF relies on both the correctness of the super-0-level set, i.e., $\mathcal{D}\subseteq\mathcal{X}_{S}$ and the existence of the control signal $\ci$ satisfying the SNCBF condition, given as follows.  
\begin{definition}[Valid Smooth SNCBFs]
\label{def:valid_smooth_SNCBF}
    The function $B(\x)$ is called a valid smooth SNCBF if $B(\x)$ satisfies the \emph{Correctness} requirement of SNCBFs, i.e. $\forall \x\in \mathcal{D}$, and the \emph{Feasibility} requirement of SNCBFs i.e. $\x\in\mathcal{X}_{S}$, and there exists a stochastic smooth NCBF control $\ci$ satisfying the conditions of Proposition \ref{prop:safety-finite-time}. 
\end{definition}
These correctness and feasibility conditions must hold for every state in the super-0-level-set $\x\in\mathcal{D}$. States are referred to as correctness or feasibility counterexamples if they violate the corresponding condition. 
Since existing VITL cannot ensure the elimination of all counterexamples \cite{so2024train,zhang2023exact}, it is essential to verify SNCBF. 

\begin{wrapfigure}{r}{0.48\columnwidth}
    \centering
    \vspace{-\intextsep}
    \includegraphics[width=0.43\columnwidth]{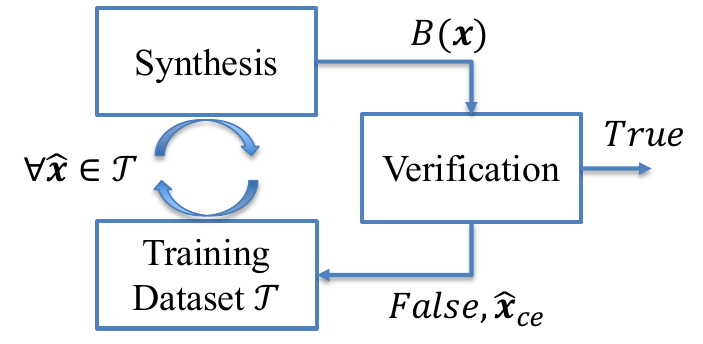}
    \caption{Workflow of the synthesis with verification in the loop.}
    \label{fig:synthesis_with_verifier}
    \vspace{-\intextsep}
\end{wrapfigure}

As shown in Fig. \ref{fig:synthesis_with_verifier}, our proposed training framework consists of three components: a training dataset $\mathcal{T}$, a synthesis module, and a verification module. These elements form two loops. The inner loop learns a parameter $\theta$ for the synthesized $B(\x)$ that satisfies the conditions of Definition \ref{def:valid_smooth_SNCBF} by minimizing a loss function over training data $\mathcal{T}$. 
The training dataset $\mathcal{T}$ is initialized through uniform sampling of $\mathcal{X}$.
This loss function consists of a weighted sum of two components, structured as an unconstrained optimization problem over $\theta$. 
\begin{equation}
    \label{eq:uncons_opt}
    \min_{\theta}\quad  \lambda_{f}\mathcal{L}_f(\mathcal{T}) + \lambda_{c}\mathcal{L}_c(\mathcal{T}) 
\end{equation}
where $\lambda_{f}$ and $\lambda_{c}$ are non-negative coefficients. The term $\mathcal{L}_f(\mathcal{T})$ is the loss penalizing the violations of constraint \eqref{eq:farkas_cons_u_1}-\eqref{eq:farkas_cons_u_3} (feasibility condition of Definition \ref{def:valid_smooth_SNCBF}), and the term $\mathcal{L}_c(\mathcal{T})$ penalizes the correctness counterexample with negative minima of \eqref{eq:smooth_containment-verification} (correctness condition of Definition \ref{def:valid_smooth_SNCBF}). 
For each sample $\hat{\x}\in \mathcal{T}$ the safe control signal $\ci(\x)$ is calculated by SNCBF-QP. The loss $\mathcal{L}_f$ enforces the satisfaction of the constraint by inserting a positive relaxation term $r$ in the constraint and minimizing $r$ with a large penalization in the objective function. We have the loss $\mathcal{L}_f$ defined as $\mathcal{L}_f = ||\ci(\x)-\ci_{ref}(\x)||^{2} + r$.
The loss term $\mathcal{L}_c$ is defined as
\begin{multline}
\label{eq:correct_loss}
       \mathcal{L}_c =  \frac{1}{N} \sum_{\hat{\x}_{i} \in \mathcal{X}_{I}} \max \left(0,\left(\epsilon -B(\hat{\x}_{i})\right) \mathds{1}_{\mathcal{X}_{s}}\right) \\
        + \frac{1}{N} \sum_{\hat{\x}_{j} \in \mathcal{X}_{U}} \max \left(0, \left(B(\hat{\x}_{j})+\epsilon  \right) \mathds{1}_{\mathcal{X}_{U}}\right). 
\end{multline}

The outer loop validates a given SNCBF $B(\x)$ by searching for counterexamples $\hat{\x}_{ce}$ and updates the training dataset as $\mathcal{T}\cup \{\hat{\x}_{ce}\}$. 
The framework penalizes constraint violations for all $\hat{\x}\in \mathcal{T}$ and extends these guarantees from $\mathcal{T}$ to the entire state space $\mathcal{X}$ through formal verification.
Utilizing this framework results in the verification of the synthesized SNCBF while also generating counterexamples to augment the training dataset.

In what follows, we present the verification of the smooth SNCBF. 
We first verify the correctness condition in Definition \ref{def:valid_smooth_SNCBF}. What we want to show is that the super-0-level set of the SNCBF is contained by the safe region $\mathcal{D}\subseteq\mathcal{X}_{S}$. The correctness counterexample refers to states $\x\in\mathcal{D}\subseteq\mathcal{X}_{S}$ with $h(\x)<0$. 
Specifically, we check if $\mathcal{D}\cap\mathcal{X}_{U}=\emptyset$. It suffices to solve the nonlinear program
\begin{equation}
\label{eq:smooth_containment-verification}
\min_{\x} \ h(\x) \quad
\mbox{s.t.} \ B(\x) \geq 0.
\end{equation}
If there exists $\x^{*}$ such that $h(\x^{*})<0$ while $B(\x^{*}) = 0$, then $\x^{*}$ is a correctness counterexample. 
Next, we verify the feasibility condition in Definition \ref{def:valid_smooth_SNCBF}. Let $\lambda(\x_{t}) := \frac{\partial B}{\partial \x_{t}} g(\x_{t})$ and 
$$\xi(\x_{t}) := \frac{\partial B}{\partial \x_{t}}f(\x_{t}) + \frac{1}{2}\mathbf{tr}(\V^{T}\frac{\partial^{2}B}{\partial \x_{t}^{2}}\V) {+} \alpha(B(\x_{t})).$$
Rearranging the terms in \eqref{eq: problem1}, we have the affine constraints given as
\begin{equation}
\label{eq:affine_cons}
{-}\lambda(\x_{t}) \ci_{t} \leq \xi(\x_{t}),
\end{equation}
The feasibility counterexample refers to the state $\x$ such that there does not exist a $\ci$ satisfying condition \eqref{eq:affine_cons}. 
We next consider the case with control input constraints $\ci\in \mathcal{U}:=\{\ci:A\ci\leq \mathbf{b}\}$. 
We present $\Lambda(\x_{t})$, $\Xi(\x_{t})$ and corresponding verification formulation. 
\begin{proposition}
\label{prop:verify_ssncbf}
Let $\Lambda(\x_{t})$ and $\Xi(\x_{t})$ be as follows. 
\begin{equation*}
\Lambda(\x_{t}) := \begin{bmatrix}
    -\lambda(\x_{t}) \\
    A
\end{bmatrix}; \qquad
\Xi(\x_{t}) := \begin{bmatrix}
    \xi(\x_{t})\\
    \mathbf{b} 
\end{bmatrix}
\end{equation*}
There always exists a $\ci$ satisfying \eqref{eq:affine_cons} if and only if there is no $\x\in\mathcal{D}$ and $\mathbf{y} \in \mathbb{R}^{n_{u}+1}$ such that
\begin{subequations}
\label{eq:farkas_cons_u}
    \begin{align}
    \label{eq:farkas_cons_u_1}
    [\mathbf{y}]_{i}\geq & 0, \ \forall i\in \{1, \ldots, n_{u}+1\} \\
    \label{eq:farkas_cons_u_2}
    -\mathbf{y}^{T}\Xi(\x) < & 0\\
    \label{eq:farkas_cons_u_3}
    \left[\mathbf{y}^{T}\Lambda(\x) \right]_{i} = & 0, \ \forall i\in \{1, \ldots, n_{u}+1\} 
\end{align}
\end{subequations}

\end{proposition}
\begin{proof}
The proof is to show that there does not exist $\x$ such that, for any $\ci \in \mathcal{U}$, \eqref{eq:affine_cons} fails to hold.  
By Farkas's Lemma, the existence of control input $\ci$ satisfying \eqref{eq:affine_cons} is equivalent to the non-existence of $\mathbf{y}$ satisfying \eqref{eq:farkas_cons_u_1}--\eqref{eq:farkas_cons_u_3}, completing the proof.
\end{proof}

The NN is a polynomial if all the activation functions are polynomials, e.g., Hermite Polynomial Activation Functions \cite{ma2005constructive}. The following lemma shows that given a polynomial SNCBF, the non-existence of $\x$ and $\mathbf{y}$ can be proved through Positivstellensatz. 
\begin{lemma}
\label{lemma:Psatz_RSNCBF}
There is no feasibility counterexample $\x\in\mathcal{D}$ if and only if there exist polynomials $\rho^{\mathbf{y}}_{1}(\x)\ldots\rho^{\mathbf{y}}_{n_{u}+1}(\x)$, sum-of-squares polynomials $q_{P}(\x)$, integers $\tau=n_{u}+3$, $r_{1}$ such that
\begin{equation}
    \label{eq:verify_ssncbf}
    \phi(\x, \mathbf{y}) + \chi(\x, \mathbf{y}) + \Gamma(\x, \mathbf{y}) = 0,
\end{equation}
\begin{align*}
    \phi(\x, \mathbf{y}) &= \sum_{K\subseteq \{1,\ldots,\tau\}} p_{i}(\x)\prod_{i\in K}\phi_i(\x, \mathbf{y})  \\
    \vspace{-0.4em}
    \chi(\x, \mathbf{y}) &=  \phi_1(\x, \mathbf{y})^{2 r_1}, \quad
    \Gamma(\x, \mathbf{y})  = \sum_{i=1}^{n_{u}}\rho^{\mathbf{y}}_{i}\left[\mathbf{y}^{T}\Lambda(\x) \right]_{i},
    \vspace{-0.4em}
\end{align*}
where $\phi_1(\cdot) = -\mathbf{y}^{T}\Xi(\x)$, $\phi_2(\cdot)=B(\x)$, and $\phi_{3,\ldots,n_{u}+3}(\cdot)=y_{i}$.
\end{lemma}
\begin{proof}
By Proposition \ref{prop:verify_ssncbf}, we have that there is no feasibility counterexample iff there exists no $\x\in\mathcal{D}$ and $\mathbf{y}$ such that \eqref{eq:farkas_cons_u_1}--\eqref{eq:farkas_cons_u_3} hold. The condition $\mathbf{y}^{T}\Xi(\x)<0$ is equivalent to $-\mathbf{y}^{T}\Xi(\x) \geq 0$ and $\mathbf{y}^{T}\Xi(\x)\neq 0$. The result then follows from the Positivstellensatz  
\end{proof}

If the SNCBF has non-polynomial activation functions, 
the conditions of Proposition \ref{prop:verify_ssncbf} can be verified by solving 
\begin{equation}
\label{eq:smooth-nonlinear-prog}
\begin{array}{ll} 
\min_{\x,y} & \mathbf{y}^{T}\Xi(\x) \\
\mbox{s.t.} & B(\x) \geq 0, \quad [\mathbf{y}]_{i}\geq 0, \ \forall i\in \{1, \ldots, n_{u}+1\} \\
& \left[\mathbf{y}^{T}\Lambda(\x) \right]_{i} = 0, \ \forall i\in \{1, \ldots, n_{u}+1\} 
\end{array}
\end{equation}
and checking whether the optimal value is nonnegative (safe) or negative (unsafe).
The following corollary describes the special case where there are no constraints on the control, i.e., $\mathcal{U} = \mathbb{R}^{n_{u}}$. 
\begin{corollary}
\label{corollary:verify_smooth_sncbf}
There exists $\ci\in \mathbb{R}^{n_{u}}$ satisfying $-\lambda(\x_{t})\ci\leq \xi(\x_{t})$
for all $\x\in\mathcal{D}$ if and only if there is no $\x$ such that
$\lambda(\x) = \mathbf{0}$, and $\xi(\x) {< 0}$.
\end{corollary}
\begin{proof}
When the vector $\frac{\partial B}{\partial x}g(\x)\neq \mathbf{0}$, there is always $u\in \mathbb{R}^{n_{u}}$ that satisfies \eqref{eq:affine_cons}. 
When the vector $\frac{\partial B}{\partial x}g(\x)= \mathbf{0}$, $\ci$ does not affect the inequality, which is equivalent to having $\ci=\mathbf{0}$. In this case, $\xi(\x_{t}) {\geq 0}$ ensures \eqref{eq:affine_cons} {holds}.
\end{proof}
Under the condition of Corollary \ref{corollary:verify_smooth_sncbf}, (\ref{eq:smooth-nonlinear-prog}) reduces to 
\begin{equation}
\label{eq:smooth-nonlinear-prog-special-case}
\min_{\x} \ \xi(\x) \quad
\mbox{s.t.} \ B(\x) \geq 0, \quad \left[\lambda(\x) \right]_{i} = 0, \ \forall i\in \{1, \ldots, n_{u}\} 
\end{equation}
The condition of Corollary \ref{corollary:verify_smooth_sncbf} can be verified by checking whether the optimal value is nonnegative (safe) or negative (unsafe).
Nonlinear programs \eqref{eq:smooth_containment-verification}, \eqref{eq:smooth-nonlinear-prog}, and \eqref{eq:smooth-nonlinear-prog-special-case} can be solved by SMT solvers with numerical completeness guarantees.

\section{Experiments}
\label{sec:experiments}

We evaluate the verifiable synthesis of a smooth SNCBF on the inverted pendulum and the unicycle model, and compare against a state-of-the-art baseline on the unicycle model.

\vspace{-0.8em}
\subsection{Experiment Settings}
The learning experiments of the SNCBFs are conducted on a computing platform equipped with an AMD Ryzen Threadripper PRO 5955WX CPU, 128GB RAM, and two NVIDIA GeForce RTX 4090 GPUs. 
We implement the networks in PyTorch, train with SGD, and verify with dReal; our code is available at \url{https://github.com/HongchaoZhang-HZ/StochasticNCBF}. Training samples states uniformly from $\mathcal{X}$ (the physical operating region of each system) without deploying the system or collecting trajectories. At deployment, the SNCBF-QP controller keeps trajectories initialized in $\mathcal{X}_I \subseteq \mathcal{X}_S$ within $\mathcal{X}_S$ with high probability; guarantees hold only while the state remains in $\mathcal{X}$.
Next, we present the inverted pendulum and the unicycle model and detailed settings.

\subsubsection{Inverted Pendulum}
We consider a continuous time stochastic inverted pendulum dynamics given as follows:
\begin{equation}
    d\left[\begin{array}{l}
    \theta \\
    \dot{\theta}
    \end{array}\right]=\left(\left[\begin{array}{c}
    \dot{\theta} \\
    \frac{g}{l}\sin(\theta)
    \end{array}\right]+\left[\begin{array}{c}
    0 \\
    \frac{1}{m l^{2}} 
    \end{array}\right] u\right) dt + \V \ d \vt,
\end{equation}
where $\theta \in \mathbb{R}$ denotes the angle, $\dot{\theta} \in \mathbb{R}$ denotes the angular velocity, $u \in \mathbb{R}$ denotes the controller, $m$ denotes the mass and $l$ denotes the length of the pendulum. We let the mass $m= 1$kg, length $l=10$m and the disturbance $\sigma = \mathsf{diag}(0.1, 0.1)$. 
The state space, initial safe region, safe region, and unsafe region are given as $\mathcal{X}=[-\frac{\pi}{4}, \frac{\pi}{4}]^{2}$, $\mathcal{X}_{I}=[-\frac{\pi}{15}, \frac{\pi}{15}]^{2}$, $\mathcal{X}_{S}=[-\frac{\pi}{6}, \frac{\pi}{6}]^{2}$, and $\mathcal{X}_{U}=\mathcal{X} \setminus \mathcal{X}_{S}$.

\iffalse %
\subsubsection{Darboux}
We consider the Darboux system~\cite{zeng2016darboux}, a nonlinear open-loop polynomial system

The dynamic model of Darboux is given as follows. 
\begin{equation}
    d\left[\begin{array}{c}
    x_1 \\
    x_2
    \end{array}\right]=\left[\begin{array}{c}
    x_2+2 x_1 x_2 \\
    -x_1+2 x_1^2-x_2^2
    \end{array}\right]dt + \V \ d \vt.
\end{equation}
We define state space, initial region, and safe region as follows. 
\begin{equation*}
    \begin{aligned}
    \mathcal{X}&=\left\{\x\in\mathbb{R}^{2}:\x \in \left[-2, 2 \right]^{2} \right\} \\
    \mathcal{X}_{I}&=\left\{\x\in\mathbb{R}^{2}:\x \in \left[0, 1\right]\times \left[1, 2\right]\right\} \\
    \mathcal{X}_{S}&=\left\{\x \in \mathbb{R}^2: x_1+x_2^2 \geq 0\right\}\\
    \mathcal{X}_{U}&=\left\{\x\in\mathbb{R}^{2}:\x \in \mathcal{X} \backslash \mathcal{X}_{S} \right\}
    \end{aligned}
\end{equation*}
\fi %

\subsubsection{Unicycle Model}

We evaluate our proposed method on a controlled system~\cite{barry2012safety}. We consider vehicles (UAVs) to avoid a pedestrian on the road. The system state consists of a 2-D position and aircraft yaw rate $\x:=[x_1, x_2, \psi]^T$. The system is manipulated by the yaw rate input $\ci$. 
\begin{equation}
    d\left[\begin{array}{l}
    x_1 \\
    x_2 \\
    \psi
    \end{array}\right]=\left(\left[\begin{array}{c}
    v\cos\psi \\
    v\sin\psi \\
    0
    \end{array}\right]+\left[\begin{array}{c}
    0 \\
    0 \\
    1 
    \end{array}\right] u\right) dt + \V \ d \vt,
\end{equation}
where $[x_p, y_p, \Theta]^T\in\mathcal{X}\subseteq\mathbb{R}^3$ is the state consisting of the location $(x_p,y_p)$ of the robot and its orientation $\Theta$ and $u_1, u_2$ controlling its speed and  orientation. 
We define the state space, initial region, safe region, and unsafe region as $\mathcal{X}=[-2, 2]^{3}$, $\mathcal{X}_{I}=[-0.1, 0.1]\times[-2, -1.8]\times[-\frac{\pi}{6}, \frac{\pi}{6}]$, $\mathcal{X}_{U}=[-0.2, 0.2]^2\times[-2, 2]$, and $\mathcal{X}_{S}=\mathcal{X}\setminus\mathcal{X}_{U}$.
\vspace{-1.5em}
\subsection{Experiment Results}
We now present the results of our smooth SNCBF experiments, reporting the hyper-parameters, settings, and outcomes for each case, and compare against the fault-tolerant SNCBF~\cite{zhang2024fault} baseline.

\subsubsection{Smooth SNCBF Evaluation}
We let $\alpha(B) = \gamma B$, where $\gamma=1$. 
We train the SNCBF assuming knowledge of the model. The SNCBF $B$ consists of one hidden layer of $20$ neurons, with Softplus activation function ($log(1+exp(\x)$). 

\begin{figure*}[htp]
\centering
\begin{subfigure}{.268\textwidth}
    \centering
    \includegraphics[width=\textwidth]{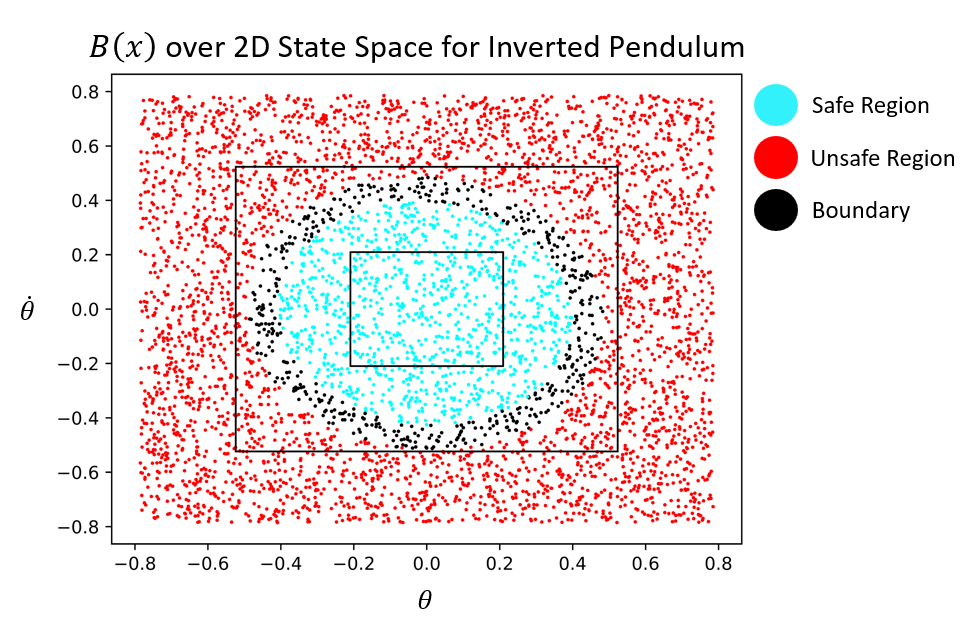}
    \subcaption{}
    \label{fig:2d_sample}
\end{subfigure}%
\begin{subfigure}{.264\textwidth}
    \centering
    \includegraphics[width=\textwidth]{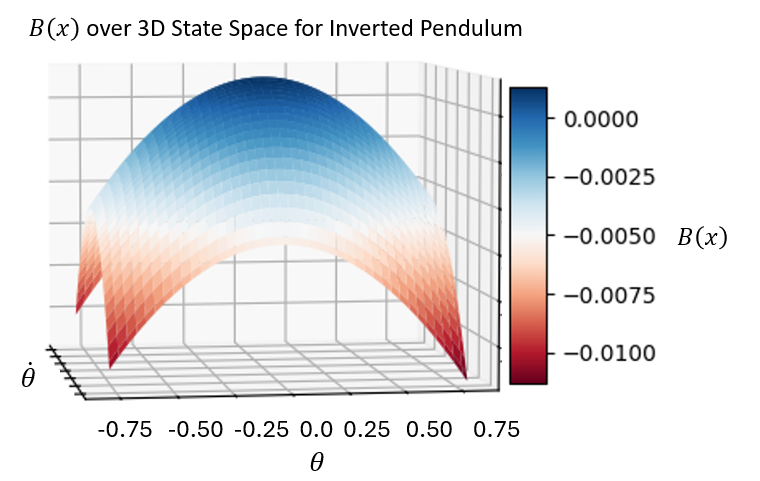}
    \subcaption{}
    \label{fig:3d_sncbf}
\end{subfigure}%
\begin{subfigure}{.22\textwidth}
    \centering
    \includegraphics[width=\textwidth]{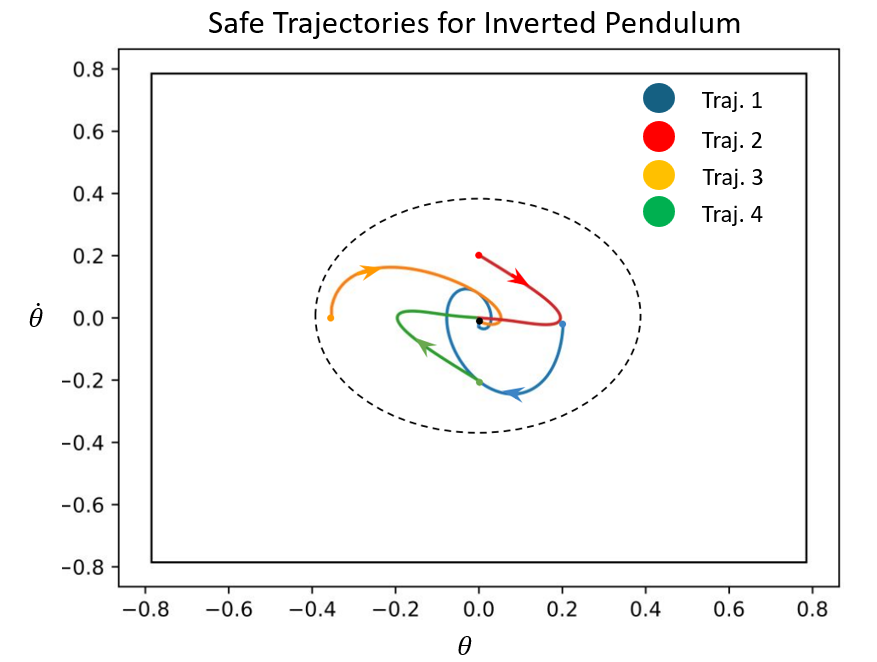}
    \subcaption{}
    \label{fig:ip_traj}
\end{subfigure}
\vspace{-0.3em}
\caption{Experimental results on the inverted pendulum. (a) $B(\x)$ over $\mathcal{X}$: blue and red denote the safe ($B \geq -\psi^*$) and unsafe ($B < \psi^*$) regions; black boxes mark the initial safe and unsafe boundaries, and the trained SNCBF boundary (black dots) separates them. (b) 3D plot of $B(\x)$ over $\mathcal{X}$. (c) Trajectories initialized in the safe set under SNCBF-QP with different reference controllers.}
\end{figure*}

\textbf{Inverted Pendulum: }
We set $\bar{\epsilon}=0.00016$, $L_{h} = 0.01$, $L_{dh} = 0.4$, and $L_{d2h} = 2$, yielding $L_{\max} = 2.4$. Training converged to $\psi^{*}=-0.00042$, so by Theorem \ref{thm: smooth_ncbf} the SNCBF is valid. Fig.~\ref{fig:2d_sample}--\ref{fig:ip_traj} visualize the trained SNCBF; validated in PyBullet, the SNCBF-QP controller keeps trajectories initialized in the safe set within it.

\begin{figure}[t]
    \centering
\vspace{-0.6em}
\includegraphics[width=0.65\linewidth]{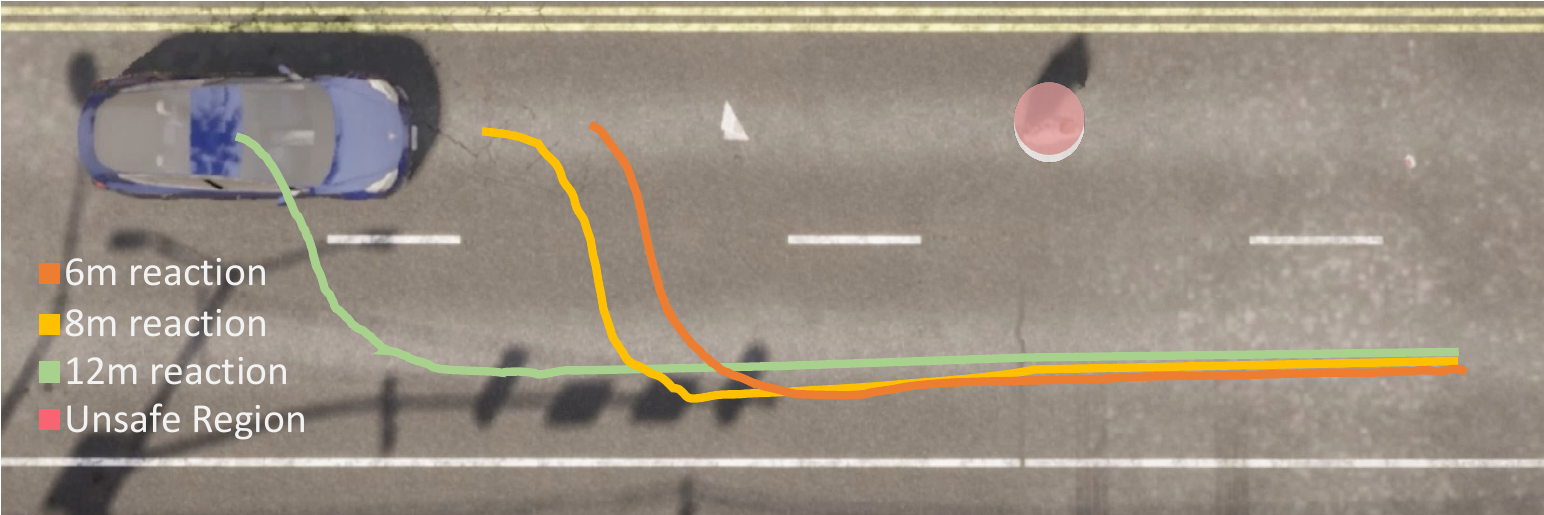}
\vspace{-0.3em}
    \caption{Vehicle trajectories under the SNCBF-QP controller at reaction distances of $6$, $8$, and $12$\,m from the pedestrian; all successfully maneuver to avoid the pedestrian.}
    \label{fig:carla_obs_avoid}
\end{figure}

\textbf{Unicycle Model: }
We validate the smooth SNCBF-based controller in the CARLA simulator, where the vehicle steers to avoid the pedestrian. We set $\bar{\epsilon}=0.01$, $L_{h} = 1$, $L_{dh} = 1$, and $L_{d2h} = 2$, yielding $L_{\max} = 4$, and training converged to $\psi^{*}=-0.04002$. Fig.~\ref{fig:carla_obs_avoid} shows the resulting trajectories at reaction distances of $6$, $8$, and $12$\,m, all avoiding the pedestrian.

%

\subsubsection{SNCBF Comparison}
\begin{table}[]
\centering
\begin{tabular}{ccc}
\toprule
\multicolumn{1}{l}{} & \begin{tabular}[c]{@{}c@{}}Baseline\end{tabular} & \begin{tabular}[c]{@{}c@{}}Smooth SNCBF\end{tabular} \\
\midrule
Verifiability    & No           & Yes        \\
Coverage         & $14.8\%$     & $72.5\%$   \\
Training Time    & $39.4$ minutes  & $64.2$ minutes   \\
\bottomrule
\end{tabular}
\vspace{-0.3em}
\caption{Comparison of baseline and smooth SNCBF synthesis proposed in Algorithm \ref{alg:sncbf}.}
\label{tab:SNCBFcomparison}
\end{table}

We compare our proposed method with an SNCBF trained with the baseline proposed in \cite{zhang2024fault} in three aspects, verifiable safety, coverage of the safe region, and the training time. The comparison results are summarized in Table \ref{tab:SNCBFcomparison}.

We observe that all synthesized SNCBFs successfully separate the safe and the unsafe regions. The baseline Fault-Tolerant SNCBF (FT-SNCBF) method did not guarantee verifiable results and achieved only $14.8\%$ coverage after training for $39.4$ minutes.
The proposed verifiable safe synthesis for smooth SNCBF achieves superior safe region coverage ($72.5\%$) with longer training time ($64.2$ minutes).

\subsubsection{Sensitivity Analysis}
To assess the impact of sampling density on SNCBF synthesis, we conducted an experiment on the unicycle benchmark varying the sample count $N$ and measuring: (i) the optimal barrier-certificate bound $\psi^*$, (ii) synthesis time $T_s$, and (iii) safe-region coverage $r_c$. Table~\ref{tab:sensitivity_N_eps_unicycle} presents the results. Recall that the validity condition (Theorem~\ref{thm: smooth_ncbf}) requires $L_{\max}\bar{\epsilon} + \psi^{*} \le 0$, where $\bar{\epsilon}$ is the radius of balls covering the domain.
\vspace{-0.3em}
\begin{table}[h]
\centering
\begin{tabular}{cccccc}
\toprule
  $N$ & $5.4\times10^{7}$ & $2\times10^{6}$ & $2.5\times10^{5}$ & $3.12\times10^{4}$ & $2\times10^{3}$ \\
  \midrule
  $\psi$ & $-0.04$ & $-0.13$ & $-0.25$ & $-0.49$ & $-1.20$ \\
  $T_s$ & 64.2 & 26.5 & 17.3 & 12.1 & 7.25 \\
  $r_c$ & 78.65\% & 78.20\% & 76.53\% & 76.72\% & 72.66\% \\
\bottomrule
\end{tabular}
\vspace{-0.3em}
\caption{Sensitivity analysis of sampling budget $N$ (and implicitly $\bar{\epsilon}$) on the unicycle system.}
\label{tab:sensitivity_N_eps_unicycle}
\end{table}
\vspace{-0.3em}
The results reveal a smooth trade-off: increasing $N$ (decreasing $\bar{\epsilon}$) yields tighter certificates (smaller $|\psi^*|$) and marginally improved coverage, but increases computation time. Notably, coverage remains stable between 72\% and 79\% across more than four orders of magnitude variation in $N$, demonstrating robustness to sampling density. A moderate sampling budget ($N \approx 10^{5}$) provides a favorable balance between synthesis time and coverage.

\vspace{-1em}
\section{Conclusion}
\label{sec:conclusion}
{We presented two complementary approaches for SNCBF synthesis with smooth activation functions. We first presented the verification-free synthesis (Section \ref{subsec:verifiable_synth}). It provides guarantees under a dense sampling assumption. To relax the assumption and to handle multiple hidden layers, we proposed the synthesis of SNCBFs with SMT-based verification in the loop (Section \ref{subsec:smooth_SNCBF_veri}).
We validated our approach on the inverted pendulum and the unicycle model. The experiments illustrated the improvement in efficiency and coverage compared to the baseline method.}

\textbf{Limitations:} 
{The verification-free approach of Section \ref{subsec:verifiable_synth} relies on dense sampling and is limited to single hidden layers due to the difficulty of computing tight global Lipschitz bounds for deep networks. The approach presented in Section \ref{subsec:smooth_SNCBF_veri} handles multiple hidden layers but with increased computational cost.}
{Scalability to higher-dimensional systems and larger networks remains a challenge. Future work will explore more efficient verification algorithms to enhance scalability and extend verification to non-smooth (e.g., ReLU) SNCBFs.}

\bibliographystyle{IEEEtran}{}
\bibliography{ref.bib}

\end{document}